# The meta-problem and the transfer of knowledge between theories of consciousness: a software engineer's take

Marcel Kvassay

Institute of Informatics, Slovak Academy of Sciences,

Dúbravská cesta 9, 845 07 Bratislava 45, Slovak Republic

marcel.kvassay@savba.sk

**Abstract**

This contribution examines two radically different explanations of our phenomenal intuitions, one reductive and one strongly non-reductive, and identifies two germane ideas that could benefit many other theories of consciousness. Firstly, the ability of sophisticated agent architectures with a purely physical implementation to support certain functional forms of qualia or proto-qualia appears to entail the possibility of machine consciousness with qualia, not only for reductive theories but also for the nonreductive ones that regard consciousness as ubiquitous in Nature. Secondly, analysis of introspective psychological material seems to hint that, under the threshold of our ordinary waking awareness, there exist further 'submerged' or 'subliminal' layers of consciousness which constitute a hidden foundation and support and another source of our phenomenal intuitions. These 'submerged' layers might help explain certain puzzling phenomena concerning subliminal perception, such as the apparently 'unconscious' multisensory integration and learning of subliminal stimuli.

As a researcher in intelligent technologies, I have long been interested in scholarly debates about consciousness. Nevertheless, my lack of formal training in philosophy would have kept me away from the arena were it not for encouragement from unexpected quarters. In this contribution, I examine two radically different explanations of our phenomenal intuitions and identify some opportunities for the transfer of knowledge between various theories. The possibility of such a mutually beneficial interaction seems to be one of the reasons David Chalmers chose the meta-problem as the focus of this symposium.

# 1. Reductive explanation

My 'reductive' choice is *Virtual machines and consciousness* (Sloman and Chrisley 2003), which is the source of all quotations in this section.

Its authors maintain that 'although the word 'consciousness' has no well-defined meaning, it is used to refer to aspects of human and animal information-processing.' They believe that a study of extant 'biological information-processing architectures' would help us supplant our vague pre-theoretical notions of consciousness with more precise and empirically tractable ones. In effect, they adopt a 'designer stance', which leads them to virtual machine functionalism (VMF).

VMF dispenses with the notion of atomic states and clean transitions between them, which shields it from 'a number of standard objections.' In VMF, a modelled entity can have several coexisting, independently-varying, and interacting states which need not start or end at the same time. VMF can thus easily accommodate the way we normally view our own mental states, such as our conflicting desires or attitudes.

The authors' first tangible result was the 'CogAff' (Cognitive-Affective) architecture schema (a 'grammar' for agent architectures) with three layers: reactive, deliberative and reflective. The reactive layer consists of the oldest and simplest mechanisms, lacking the ability to explicitly represent and compare alternatives in a symbolic form. Consequently, it cannot 'reason about nonexistent or unperceived phenomena (e.g., future possible actions or hidden objects)', which is the core ability of the second, deliberative layer. The third, reflective layer (also called meta-management) contains reflective and self-reflective mechanisms, whose main task is to monitor deliberative processes and, should they get stuck, interrupt them and redirect the processing to more promising alternatives.

On this basis, the authors formulate their 'human-like architecture for cognition and affect' ('H-CogAff'), which they think sufficient (in principle) for the construction of robots with human-like common sense and consciousness. Their claim rests on an 'unrestricted' form of VMF, in which states and processes need not be causally connected to the rest of the system or its inputs and outputs. They maintain that unrestricted VMF can explain even notoriously elusive features of human psychology such as *qualia*. They tried to demonstrate this through a graded series of examples, which culminated in a special case of a semi-detached process in the reflective layer that monitors and evaluates other processes:

> *This internal self-observation process might have no causal links to external motors, so that its information cannot be externally reported. If it also modifies the processes it observes ... then it may have external effects. However it could be the case that the internal monitoring states are too complex and change too rapidly to be fully reflected in any externally detectable behaviour: a bandwidth limitation. For such a system experience might be partly ineffable. (p. 151)*

The authors seem to imply that partial ineffability in the reflective layer amounts to a kind of proto-qualia, or elementary qualia, instantiated in a machine. They bolster their argument through a discussion of concept formation in self-organising systems, which, as they show, provide scope for more advanced forms of qualia. To that end, they draw a distinction between 'architecture-based' and 'architecture-driven' concepts. While the architecture-based concepts are defined 'in terms of what components of the architecture can do', and refer to their states and processes as if 'from outside', the architecture-driven ones are inherently private, especially if they form during the unique developmental histories of such systems without any reference to an external world. Such concepts then acquire 'causal indexicality' (see Campbell 1994):

> *If such a concept C is applied by A to one of its internal states, then the only way C can have meaning for A is in relation to the set of concepts of which it is a member, which in turn derives only from the history of the self-organising process in A.... This can be contrasted with what happens when A interacts with other agents in such a way as to develop a common language for referring to features of external objects. Thus A could use 'red' either as expressing a private, causally indexical, concept referring to features of A's own virtual-machine states, or as expressing a shared concept referring to a visible property of the surfaces of objects. (p. 165)*

Considering two agents, A and B, of this type,

> *then if A uses its causally indexical concept $C_a$, to think the thought 'I am having experience $C_a$', and B uses its causally indexical concept $C_b$, to think the thought 'I am having experience $C_b$', the two thoughts are intrinsically private and incommunicable, even if A and B actually have exactly the same architecture and have had identical histories leading to the formation of structurally identical sets of concepts. A can wonder: 'Does B have an experience described by a concept related to B as my concept $C_a$ is related to me?' But A cannot wonder 'Does B have experiences of type $C_a$', for it makes no sense for the concept $C_a$ to be applied outside the context for which it was developed, namely one in which A's internal sensors classify internal states. They cannot classify states of B. (p. 165)*

Such self-referential architecture-driven concepts, the authors claim,

> *are strictly non-comparable: not only can you not know whether your concepts are the same as mine, the question is* incoherent. *If we use the word 'qualia' to refer to the virtual machine states or entities to which these concepts are applied, then asking whether the qualia in two experiencers are the same would be analogous to asking whether two spatial locations in different frames of reference are the same, when the frames are moving relative to each other. But it is hard to convince some people that this makes no sense, because the question is grammatically well-formed. Sometimes real nonsense is not* obvious *nonsense. (p. 166)*

The authors thus explain the spontaneous emergence of 'phenomenal intuitions' about qualia in conscious creatures of the requisite sophistication by their internal structure: 'Some robots with our information processing architecture,' they conclude, 'will discover qualia and be puzzled about them. The more intelligent ones should accept our explanation of how that happens.'

In a sequel (Chrisley & Sloman 2016), they classify themselves as 'qualia revisionists' and admit that qualia do (or might) exist, with the important proviso that they may not be as they seem. This differentiates them from both qualia eliminativists, like Daniel Dennett, and 'naive qualia realists', who hold that qualia exist and are truly as they seem: unmediated, private, intrinsic, and ineffable.

Although I subscribe to strongly nonreductive views on consciousness, the authors have convinced me that intelligent machines could possess qualia. More precisely, I feel we should rather speak of proto-qualia, structural and functional preconditions leading to qualia in the presence of some form of irreducible consciousness. Such a possibility for machines is inherent e.g. in Chalmers' minimalistic 'nonreductive functionalism' (Chalmers 1995). In principle, therefore, its proponents could directly incorporate Sloman's and Chrisley's conceptions into their theories, thereby strengthening the case for the feasibility of human-like machine consciousness. (For a more detailed comparison of Chalmers with Sloman and Chrisley see Kvassay 2012). However, it might still turn out that there are kinds of qualia that only humans can have but that would not detract from the value of demonstrating the viability of machine consciousness with qualia on a nonreductive basis, given that people with strongly nonreductive views tend to resist its very idea.

## 2. Nonreductive explanation

Chalmers' nonreductive functionalism represents an intermediate position between physicalism and strongly nonreductive views, which is an ideal vantage-point for initiating a meaningful dialogue between them. Chalmers himself tentatively explores the potential of idealism to solve the mind-body problem in *Idealism and the Mind-Body Problem* (forthcoming). In doing so, he differentiates its numerous varieties, but there is one important distinction that he omits: the one between purely speculative systems and those associated with psychological practices such as concentration and meditation, which also attempt to experience the higher or deeper states of consciousness postulated by these systems. I believe that their accumulated psychological material could help explain many aspects of the problem of consciousness and the meta-problem, if approached from the right angle and with an adequate explanatory framework. Naturally, to develop such a framework would not be easy because each system tends to cast its psychological experiences into a form congruent with its underlying philosophy. Many are openly theistic, though some are non-theistic or even atheistic (see, e.g. Heehs 2014, 2018). Nevertheless, a competent multidisciplinary team might still achieve worthwhile results.

In order to illustrate the idea, I will now review some salient observations by the Indian thinker Sri Aurobindo (1872-1950), whose main philosophical work *The Life Divine* (Aurobindo 2005) can be regarded as one such attempt. In this tome, he developed his own position ('a universal Realism') and contrasted it with Materialism and 'spiritual Illusionism'. In the process, he reflected on the rich experiential material drawn mainly from Indian Vedanta and revaluated it in the light of modern knowledge and his own spiritual experiences. I will try to show that the result is highly relevant for the contemporary philosophy of mind, including the meta-problem. (All quotations in this section are from *The Life Divine*.)

Arguably the most pertinent are his remarks on the function and constitution of the human mind, particularly in Book Two, chapters 8-11 (pp. 519-85). Many are philosophically neutral but are embedded in his idealistic metaphysics, from which they have to be extricated. Thus, for example, he notes that 'Ordinarily, we speak of a subconscious existence and include in this term all that is not on the waking surface.' We are inclined to think of it as something devoid of consciousness and inferior to our waking mind. To him, this is an error. He calls that large submerged part of us our 'subliminal' self and distinguishes in it four distinct layers: (1) the subconscious, (2) the submental, (3) the inner or subliminal proper and (4) the superconscient or supramental. Of these, only the first two are really inferior, although still *not* completely devoid of consciousness; the third is qualitatively similar, yet much more capable than our waking mind; and the last is so much superior that it is practically inaccessible to all but the most advanced spiritual practitioners. In their totality, they represent 'our real or whole being, of which the outer [waking consciousness] is a part and a phenomenon, a selective formation for a surface use' (p. 576).

Our submental self corresponds to 'a vitality working in [our] bodily form and structure as in the plant or lower animal.' Because it is mostly subconscious to us, we tend to think that 'this vital-physical part of us also is not conscious of its own operations' and 'becomes conscious only so far as it is enlightened by mind and observable by intelligence.' But this is a mistake, he writes,

> *due to our identification of consciousness with mentality and mental awareness. Mind identifies itself to a certain extent with the movements proper to physical life and body and annexes them to its mentality, so that all consciousness seems to us to be mental. But if we draw back, if we separate the mind as witness from these parts of us, we can discover that life and body… have a consciousness of their own, a consciousness proper to an obscurer vital and to a bodily being, even such an elemental awareness as primitive animal forms may have, but in us partly taken up by the mind and to that extent mentalised. (p. 579)*

He admits that it "has not, in its independent motion, the mental awareness which we enjoy' because 'there is no organised self-consciousness,' yet 'it has its own separate reactions to contacts and is sensitive to them in its own power of feeling; it does not depend for that on the mind's perception and response.' The true subconscious, he explains next,

> *is other than this vital or physical substratum; it is the Inconscient vibrating on the borders of consciousness, sending up its motions to be changed into conscious stuff, swallowing into its depths impressions of past experience as seeds of unconscious habit and returning them constantly but often chaotically to the surface consciousness… in dream, in mechanical repetitions of all kinds, in untraceable impulses and motives,… in dumb automatic necessities of our obscurest parts of nature. (pp. 579-80)*

Our inner self (the subliminal proper) is qualitatively different from both because

> *it is in full possession of a mind, a life-force, a clear subtle-physical sense of things. It has the same capacities as our waking being… [but] wider, more developed, more sovereign…. [I]t exceeds the physical mind and physical organs although it is aware of them and their works and is, indeed, in a large degree their cause or creator. It is only subconscious in the sense of not bringing all or most of itself to the surface. (p. 580)*

The extraordinary capabilities of this inner self underpin such apparently 'miraculous' phenomena as hypnotic analgesia and anaesthesia (p. 115). They can also manifest spontaneously in a special type of dreams, which sometimes bring us solutions to problems that we could not solve in our waking state (pp. 440-41).

The superconscient represents a yet higher level of 'extraordinariness.' It is usually associated with deeply spiritual and religious experiences, but it does have one 'secular' manifestation, too: the phenomena of genius. In the genius, though, there is a certain veiling element,

> *because the light of the superior consciousness not only acts within narrow limits, usually in a special field,… but also in entering the mind it subdues and adapts itself to mind substance so that it is only a modified or diminished dynamis that reaches us, not all the original divine luminosity of what might be called the overhead consciousness beyond us. (p. 289)*

In most people, however, the superconscient acts rarely and indirectly through their subliminal proper.

With respect to these four different components, our waking consciousness appears as a makeshift construction in a permanent state of flux, an amalgam of disparate materials and influences rising from them to our waking surface. This gives our subjective existence a very complicated (and partly chaotic) character and explains, among other things, why our attempts at rational self-control are never entirely or permanently successful. It also leads to several classes of phenomena that could be considered phenomenal intuitions.

The simplest type rests on the inability of our waking mind (a weakness shared by its subliminal counterpart) to directly perceive the indivisible unity of all existence, including matter and consciousness. In Aurobindo's system, such a unitary perception is the prerogative of the superconscient and, until our mind is 'transformed' by its direct contact and influence, our ordinary perception will continue to render matter and consciousness, as well as other pairs of dualities like pleasure/pain, personality/impersonality, activity/passivity, truth/falsehood, good/evil, etc., as irreconcilable opposites, irrespective of whether we intellectually adhere to some form of philosophical monism or dualism.

The second type builds on the first and adds a vague valuation: consciousness may be perceived as the worthier of the two because we tend to regard matter as mere inert stuff, but consciousness seems to be both the essence of ourselves and the seat of our precious capacity to fashion that material creatively in innumerable ways.

Finally, intuitions and intimations of the third type (usually mediated by our subliminal proper) make us aware of, and dissatisfied with, our present imperfect condition. They spur us on towards continual self-exceeding in the never-ending quest for a more integrated inner life and an ever-expanding fullness of being. In spiritually inclined people, this often takes the form of regular spiritual practice, but in the secularly minded, it may manifest as a thirst for knowledge, a search for the meaning of life, a self-dedication to the betterment of humanity or some other similar ideal.

Aurobindo's philosophical *magnum opus* was in fact meant for people in the third category, who felt moved by 'the impulse towards perfection, the search after pure Truth and unmixed Bliss, the sense of a secret immortality' (p. 3). Today we might prefer to put it differently but, verbal differences aside, do we not still continue to dream of perfect knowledge, unlimited happiness, unshakable health, and perhaps even of the conquest of death through the 'miracles' of science and technology?

Of course, in materialist theories, all the above types of phenomenal intuitions would be replaced by their 'computational' or 'functional' equivalents, with appropriate evolutionary justifications. Thus, for example, the need to differentiate between pleasant and painful, beneficent and maleficent is a clear precondition for survival. Aurobindo actually admits this when he calls pain a 'device of Nature' (p. 115), but he still needs a better alternative compatible with unitary consciousness. Analogously, the second type might simply be a clever trick of Nature, reinforcing our instinct of self-preservation; and the third, a similar device, predisposing us towards change instead of being attached to one static optimum (which might turn into an evolutionary liability). I will not attempt to adjudicate these issues here, nor will I, for reasons of space limitations, discuss in further detail Aurobindo's philosophy or his acute introspective insights. A handy summary of the former can be found in Odin (1981), and of the latter in Kvassay (2011).

To summarise, although Aurobindo's idealistic metaphysics is unlikely to find many takers among contemporary philosophers, his concept of several 'subliminal' layers of consciousness might have a broader appeal. In fact, something similar seems to be implied by Sloman's and Chrisley's CogAff schema, because each of its layers can be construed as a distinct and independently operating type of consciousness. For example, if we grant purely reactive species like termites a rudimentary awareness ('submental' in Aurobindo's terminology), then this type of awareness might survive in a 'submerged' form (perhaps with some loss of priority and importance), even as new conscious layers (deliberative and reflective) are added to that reactive foundation in its evolutionary descendants and inheritors. Moreover, Sloman himself speaks of 'two co-occurrent forms of consciousness, one performing a task and succeeding or failing, the other observing modal aspects of task e.g.

necessity or impossibility' (personal communication). I believe this 'confluence' of views testifies to the value of the idea. It might also help to explain certain puzzling phenomena regarding subliminal perception. In a recent study in which Chrisley himself participated (Scott et al. 2018), associative learning occurred between pairs of subliminal stimuli presented in different sensory modalities, thus challenging theories of consciousness based on the 'conscious access hypothesis' (CAH). The theoretical admission of a concurrently active submental layer of consciousness in the experiment participants (of which they would not have been directly aware) could explain the results without completely undermining the CAH, although its revision would probably still be needed.

# Conclusion

In this paper, I have examined two different approaches to consciousness which, for various reasons, seem to have stayed largely beneath the radar of most professional philosophers of mind. Sloman's and Chrisley's pragmatic approach, perhaps a bit too technical and empirical at first sight, may yet be essential for the success of the more abstract philosophical enterprise; first, by helping it to clarify inherently vague concepts like 'consciousness' and, second, by practically demonstrating the capabilities of sophisticated agent architectures, such as their support for certain functional forms of qualia or proto-qualia. As a software engineer, I am in full sympathy with their 'designer stance' and hope to see more practical results coming out of their continual explorations of new territories, such as learning mechanisms in 'altricial' versus 'precocial' animals, the role of mathematical intuition versus mathematical ingenuity, or the need for chemistry-based forms of computation (Sloman 2013, 2018).

Aurobindo's ideas seem to have been neglected for opposite reasons. First, they are part of an unabashedly idealistic system with a strong Vedantic colouring, which runs counter to the broadly analytical tenor of the contemporary philosophy of mind. Second, he often employs a demanding style in which, to quote his biographer, 'clause follows clause follows clause, until sometimes the point of the statement is lost in a maze of qualifications' (Heehs 2008, p. 328). But whatever the merits and demerits of his philosophy and style, his incisive introspective insights remain universally relevant. Probably the most pertinent of these is his claim, broadly compatible with Sloman's and Chrisley's architectural notions, that, under the threshold of our waking mind, there exist several distinct and concurrently active layers of 'submerged' or 'subliminal' consciousness. These constitute an independent source of our phenomenal intuitions and factoring them into extant theories of consciousness could open the way to solving several conundrums presently facing them, such as the apparently 'unconscious' multisensory integration and learning of subliminal stimuli.

# Acknowledgments

This work was supported in part by the Slovak Research and Development Agency under contract No. APVV-17-0619 and by the VEGA project No. 2/0167/16. I am grateful to Aaron Sloman for comments on an earlier draft of this manuscript.